
\magnification=\magstep1
\baselineskip=20pt
\centerline{Delayed Hard Photons from Gamma-Ray Bursts}
\bigskip
\centerline{J. I. Katz}
\centerline{Department of Physics and McDonnell Center for the Space
Sciences}
\centerline{Washington University, St. Louis, Mo. 63130}
\centerline{I: katz@wuphys.wustl.edu}
\bigskip
\centerline{Abstract}
\medskip
The delayed hard (up to 25 GeV) photons observed more than an hour following
a gamma-ray burst on February 17, 1994 may result from the collisions of
relativistic nucleons with a dense cloud, producing $\pi^0$.  The required
cloud density is $\sim 2 \times 10^{11}$ cm$^{-3}$.  This cloud may be the
remains of the disrupted envelope of a neutron star, and may survive as an
excretion disc of $\sim 10^{14}$--$10^{15}$ cm radius around the coalescing
binary.
\vfil
\noindent
Subject headings: Gamma-rays: Bursts---Stars: Neutron
\eject
\centerline{1. Introduction}
\medskip
The recent observation by EGRET (Hurley, {\it et al.} 1994) of a $\approx
25$ GeV photon from the direction of a gamma-ray burst (GRB) on February 17,
1994, but following it by about 77 minutes, and of several $\sim 100$ MeV
photons at comparable (but not identical) time lags, cannot be accommodated
within models in which a GRB is the result of synchrotron emission in a
collisionless shock-heated plasma produced by relativistic fireball debris
impinging upon interstellar gas (Rees and M\'esz\'aros 1992, M\'esz\'aros
and Rees 1993, Katz 1994).  The time delay is the difficulty.  In these
models high energy radiation is associated with high Lorentz factors, short
pulse durations and short intervals from the onset of emission.  At later
times synchrotron radiation is observed from fireball debris whose kinetic
energy has been degraded by sweeping up interstellar matter, and its
spectrum will therefore be softer than that observed early in the burst.
Further, a burst duration of more than an hour could only be obtained for
interstellar densities {\it many} orders of magnitude less than 1 cm$^{-3}$.
In the case of most GRB (including that of February 17, 1994) this would be
inconsistent with the values of the parameters implied by their emission of
$\sim$ MeV photons over durations of seconds to minutes.

I suggest an explanation of these observations as the result of collisions
between energetic nucleons in the fireball debris and a dense cloud of low
velocity gas near the site of the GRB.  \S2 contains estimates of the
required parameters.  In \S3 I discuss the problem of forming and
maintaining the required cloud.  \S4 contains a brief summary discussion.
\bigskip
\centerline{2. Collisional Gamma-Rays}
\medskip
The observed energetic gamma-rays may be produced by the process
$$\eqalignno{p + p &\to p + p + \pi^0, &(1a)\cr
\pi^0 &\to \gamma + \gamma. &(1b)\cr}$$
A significant fraction of the nucleons in the universe are neutrons,
stabilized by their presence in helium or heavier nuclei, so that
$$p + n \to p + n + \pi^0 \eqno(1a^\prime)$$
may often take the place of (1a).  Reaction channels which include products
in addition to $\pi^0$ should be considered at high energies.

The total cross-section of nucleons on nucleons at multi-GeV energies is
roughly 30 millibarns, nearly independent of energy.  The partial
cross-section for $\pi^0$ production is several times smaller.  The
laboratory frame energies of the gamma-rays are typically about 10\% of that
of the incident nucleon, so the observation of a $\approx 25$ GeV photon
suggests a nucleon of $\sim 300$ GeV, corresponding to a Lorentz factor
$\gamma \sim 300$.  This is within the range assumed in fireball models of
GRB, and might be taken as support for those models if the hard gamma-ray
production can be explained.

The chief difficulty in fireball models of GRB is turning debris kinetic
energy into observable gamma-rays, which is why these models usually assume
collective interactions (collisionless shocks).  Without collective
interactions, at ordinary interstellar densities (1 cm$^{-3}$) the
nucleon-nucleon interaction length is about 10 Mpc!  An alternative
resolution of this problem is to assume extraordinarily high densities.
This cannot explain the lower energy emission of GRB, because the resulting
radiation is a combination of high energy (mean energy $\ge 70$ MeV)
gamma-rays from $\pi^0$ decay and visible, ultraviolet, and X-ray radiation
from the heated matter, and because the time scales are much too long, but
it may be the explanation of the delayed hard gamma-rays.

The characteristic interaction time of a relativistic nucleon moving through
a gas of nucleon density $n$ (the mean density averaged over particle paths)
is
$$t \sim {1 \over n \sigma c}, \eqno(2)$$
where $\sigma \approx 3 \times 10^{-26}$ cm$^2$ is the total interaction
cross-section.  Here we assume that the energetic particles are moving
roughly isotropically through the gas cloud (collimation toward the
observer would reduce the observed duration), so that the observed $t
\approx 5 \times 10^3$ s may be used in (2), with the result
$$n \sim 2 \times 10^{11}\ {\rm cm}^{-3}. \eqno(3)$$
The spatial extent of such a cloud of mass $M_{cl}$, described as a
homogeneous sphere of radius $r$ (surely an oversimplification) is
$$r \sim \left({M_{cl} \over M_\odot}\right)^{1/3} \left({3 \times 10^{56}
\over n}\right)^{1/3} \sim \left({M_{cl} \over M_\odot}\right)^{1/3}
\left({t \over 5 \times 10^3\ {\rm s}}\right)^{1/3} 1 \times 10^{15}\ {\rm
cm}. \eqno(4)$$
Because the observed $ct \sim 10^{14}$ cm the observed radiation comes from
only a small inner fraction of the cloud if $M_{cl} \gg 10^{-3} M_\odot$.
If $ct > r$ were observed it would not contradict the model because
isotropized particles, gyrating in a magnetic field, may spend a time much
longer than $r/c$ inside the cloud.  For a 300 GeV proton a gyroradius of
$10^{14}$ cm corresponds to $B \approx 10^{-5}$ gauss, only a few times
greater than typical interstellar values and a modest value for a cloud of
the required density.

The total column density of the cloud is
$$nr \sim \left({M_{cl} \over M_\odot}\right)^{1/3} \left({t \over 5 \times
10^3\ {\rm s}}\right)^{-2/3} 2 \times 10^{26}\ {\rm cm}^{-2}. \eqno(5)$$
The dominant source of opacity of hydrogenic matter to energetic gamma-rays
is pair production.  The cross-section per hydrogen atom, assuming complete
screening and including pair production by the electrons, is $1.7 \times
10^{-26}$ cm$^2$, essentially independent of energy for $E_\gamma > 1$ GeV
and logarithmically less at lower energies where screening is less
complete (Heitler 1954).  The implied optical depth to escape of gamma-rays
is
$$\tau_{pair} \sim 3 \left({M_{cl} \over M_{\odot}}\right)^{1/3} \left({t
\over 5 \times 10^3\ {\rm s}}\right)^{-2/3}. \eqno(6)$$
Energetic gamma-rays escape if $M_{cl} < 0.03\ M_{\odot}$.  The proton
interaction cross-section is only about twice that for absorption of the
gamma-rays, but the gamma-rays may escape for a wide range of cloud
parameters and geometries because photons follow straight line paths while
the protons may gyrate many times through a cloud of modest dimensions.

At photon energies below 50 MeV the Klein-Nishina cross-section exceeds that
of pair production, and Compton scattering is the dominant source of
opacity.  The Compton optical depth is approximately
$$\tau_{Compt} \sim 4 \left({M_{cl} \over M_{\odot}}\right)^{1/3} \left({t
\over 5 \times 10^3\ {\rm s}}\right)^{-2/3} \left({50\ {\rm MeV} \over
E_\gamma}\right) \eqno(7)$$
for $m_e c^2 \ll E_\gamma$, where the logarithmic factor has been taken as a
constant.

The threshold energy $\hbar \omega_{th}$ for $\gamma$-$\gamma$ pair production
by a gamma-ray of energy $E_\gamma$ is $\hbar \omega_{th} = (m_e c^2)^2 /
E_\gamma$, and is 10 eV for $E_\gamma = 25$ GeV and 2.6 KeV for $E_\gamma
= 100$ MeV.  The ultraviolet photon density will be very low in a cool dense
cloud, so that even energetic gamma-rays will not be attenuated by this
process.  If the cloud becomes heated (by deposition and thermalization of
the energy of the energetic nucleons) and is optically thick it will fill
with a black body radiation field at a temperature $T_{bb}$; if $k_B T_{bb}
> 0.1 \hbar \omega_{th}$ the optical depth for $\gamma$-$\gamma$ pair
production will typically become very large, and the most energetic
gamma-rays will not escape.  This will introduce an energy-dependent cutoff,
with only gamma-rays satisfying the condition
$$E_\gamma < 0.1 {(m_e c^2)^2 \over k_B T_{bb}} \eqno(8)$$
escaping.  This cutoff will also vary as the thermal radiation field
changes.  There is thus a transparency window for gamma-rays between
attenuation by $\gamma$-$\gamma$ pair production at high energies (8) and
attenuation by Compton scattering at low energies (7).
\bigskip
\centerline{3. The Cloud}
\medskip
The model proposed here depends on the existence of a dense cloud near the
source of energetic particles.  The required density is so high that the
cloud cannot be interstellar, and must be associated with the source of the
GRB.  If we accept the hypothesis (Eichler, {\it et al.}~1989) that GRB have
their origin in the coalescence of two orbiting compact objects, we should
look to the earlier stages of that coalescence as the source of the required
matter.  Degenerate dwarfs and nondegenerate stars are not possible sources
because the mass-radius relations of these objects imply that as they lose
mass their densities decrease and their orbital periods increase; the
mass-losing star would only erode slowly until it disappeared.  Accelerating
orbital evolution culminating in the cataclysmic gravitational
radiation-driven coalescence required to make a GRB will occur only if the
coalescing objects are neutron stars or black holes.

The simplest explanation of the cloud is that it is the remains of an
extended envelope around an inspiraling neutron star (a structure analogous
to that of a red supergiant star, though perhaps smaller and less massive).
Models of such objects were calculated by Thorne and \.Zytkow (1977).  I use
their calculated static structures to describe envelopes undergoing dynamic
stripping.  The characteristic inspiraling time for one neutron star
orbiting in the envelope of another is
$$t_{sp} \sim \left({M \over \rho a^3}\right) \left({a^3 \over
2GM}\right)^{1/2}, \eqno(9)$$
where $a$ is the separation between the two neutron stars, $M$ is the mass
of each, and $\rho$ is the density of the envelope through which the
intruder is passing.  In the outer layers of the model envelopes $\rho \sim
0.1 M/a^3$ and $t_{sp}$ is of order the Keplerian orbit time of several
years.  However, at smaller $a$ the calculated $\rho$ is as small as $3
\times 10^{-17} M/a^3$, and $t_{sp}$ is much larger.  For $a < 10^{10}$ cm
gravitational radiation (rather than hydrodynamic drag) is the dominant
mechanism of angular momentum loss.  The slowest stage of orbital evolution
occurs when $a \approx 10^{10}$ cm, and has a characteristic time of $\sim 3
\times 10^{12}$ s for the 5 $M_\odot$ model and $\sim 2 \times 10^{11}$
s for the 12 $M_\odot$ model of Thorne and \.Zytkow (1977).  These
estimates are crude, and the applicability of the static structures
uncertain, but they indicate that rapid disruption of the outer stellar
envelope is followed by a longer period of slow orbital decay before it is
accelerated by gravitational radiation.

The expelled matter may flow through the outer Lagrange point to form an
excretion disk whose initial radius is about twice the envelope's radius, or
$\sim 10^{14}$ cm. This is comparable to the light (or energetic particle)
travel time size implied by the delayed gamma-rays of the GRB of February
17, 1994, although the time scale may instead be explained by Eq. (2).
Matter which has escape velocity will reach distances $\sim 1$ pc by the
time the neutron stars coalesce, and will be too dilute and distant to have
any effect.  If the disrupted envelope is massive the disk may be subject to
self-gravitational instabilities, whose result is incalculable; much of it
may be lost.  The fraction which must remain when the GRB occurs is small,
and the necessary survival time is only $\sim 10^2$--$10^3$ disk orbits (at
$r \sim 10^{14}$ cm), so that it is likely to be there when needed.

Alternatively, it is possible that the slowest stage of orbital evolution
only lasts $\sim 10^8$--$10^9$ s because the disrupting envelope
restructures itself and exerts more drag on the inspiraling neutron star
than implied by static models.  In this case the cloud may be in free
expansion with $r \sim 10^{15}$ cm when the GRB occurs.

The actual geometry of the cloud must be complex.  In order to observe the
initial GRB our line of sight must be transparent to soft gamma-rays.  The
time scales of the GRB require a density closer to interstellar values than
those discussed here.  Yet a substantial fraction of the fireball debris
must intercept dense matter, or be captured on magnetic field lines which
enter it, in order to produce the delayed gamma-rays.  Very heterogeneous
distributions are required, with a dense disk or clouds and low density
regions elsewhere.  During the period of slow orbital evolution a dilute
wind from the continuing disruption of the inner envelope may blow a bubble
inside the massive disrupted envelope, perhaps out of the plane of the disk,
with low enough density to permit formation of a ``classical'' GRB with the
observed duration.
\bigskip
\centerline{4. Discussion}
\medskip
The energy radiated in delayed gamma-rays is not small compared to that in
the prompt GRB.  The usual estimate of $10^{51}$ erg for a GRB at cosmological
distances then implies a comparable energy collisionally deposited in the
dense cloud by the relativistic particles.  Such an event may qualitatively
resemble a Type II supernova, aside from its gamma-radiation.  The
pre-outburst star would show evidence of rapid mass loss and might also be a
binary X-ray source of short period (most of its evolution is spent with $a
\sim 10^{10}$ cm and an orbital period $\sim 5$ minutes).  Following the
period of rapid mass loss the expanding envelope might resemble a planetary
nebula, or fade to invisibility because of the absence of ultraviolet
excitation.  Continuing mass loss from the remains of the envelope might
prevent the observation of these systems as binary pulsars, and therefore
they might not be included in predictions of the frequency of neutron star
coalescence.

The origin of a neutron star with an envelope is also speculative, but might
be a quiet core collapse in an ordinary supergiant or the capture of an
envelope by a naked neutron star in collision with a nondegenerate star.
The latter process is plausible in dense galactic or globular cluster
nuclei, in which binary evolution is often determined by close stellar
encounters.  Collisional capture is likely to be inefficient, leading to
envelope masses $\ll M_\odot$.

I thank B. L. Dingus and K. Hurley for unpublished data, B. E. Schaefer for
discussions and NASA NAGW-2918 for support.
\vfil
\eject
\centerline{References}
\parindent=0pt
\def\ref{\medskip \hangindent=20pt \hangafter=1}
\ref
Eichler, D., Livio, M., Piran, T. and Schramm, D. 1989 Nature 340, 126
\ref
Heitler, W. 1954 {\it Quantum Theory of Radiation} 3rd ed. (Clarendon Press,
Oxford)
\ref
Hurley, K., {\it et al.} 1994 preprint
\ref
Katz, J. I. 1994 ApJ 422, 248
\ref
M\'esz\'aros, P. and Rees, M. J. 1993 ApJ 405, 278
\ref
Rees, M. J. and M\'esz\'aros, P. 1992 MNRAS 258, 41p
\ref
Thorne, K. S. and \.Zytkow, A. N. 1977 ApJ 212, 832
\vfil
\eject
\bye
\end